\documentclass[twocolumn, prb, aps, superscriptaddress,floatfix]{revtex4-1}
\usepackage{float}
\usepackage{siunitx}
\usepackage{amsmath}
\usepackage{bm}
\usepackage{siunitx}
\usepackage{xcolor}
\usepackage[colorlinks=true, allcolors=blue]{hyperref}
\usepackage{color}
\usepackage{chemformula}
\usepackage{siunitx}
\usepackage{graphicx}
\usepackage{bm}
\usepackage{color}
\usepackage[utf8]{inputenc}\usepackage[T1]{fontenc}
\usepackage{amsmath}
\usepackage{epsfig}
\usepackage{epstopdf}
\usepackage{multirow}
\usepackage{comment}
\usepackage{physics}
\usepackage{multirow}
\usepackage[colorlinks=true, allcolors=blue]{hyperref}
\usepackage{mathrsfs}
\usepackage{natmove}
\begin{document}
\title{Superconductivity of $\beta$-Gallium}
\author{Yundi Quan}
\affiliation{Department of Physics, University of Florida, Gainesville, Florida 32611, USA}
\affiliation{Department of Materials Science and  Engineering, University of Florida, Gainesville, Florida 32611, USA}
\affiliation{Quantum Theory Project, University of Florida, Gainesville, Florida 32611, USA}
\author{P.\ J.\ Hirschfeld}
\affiliation{Department of Physics, University of Florida, Gainesville, Florida 32611, USA}
\author{R.\ G.\ Hennig}
\affiliation{Department of Materials Science and  Engineering, University of Florida, Gainesville, Florida 32611, USA}
\affiliation{Quantum Theory Project, University of Florida, Gainesville, Florida 32611, USA}

\date{\today}
\begin{abstract}
Elemental gallium can exist in several phases under ambient conditions. The stable $\alpha$ phase has a superconducting transition temperature, $T_c$, of 0.9~K. By contrast, the $T_c$ of the metastable $\beta$ phase is around 6~K. To understand the significant improvement in $T_c$ in the $\beta$ phase, we first calculate the electronic structure, phonon dispersion, and the electron-phonon coupling of gallium in the $\alpha$ and $\beta$ phase. Next, we solve the Eliashberg equations to obtain the superconducting gaps and the transition temperatures. Using these results, we relate the increased $T_c$ in the $\beta$ phase to structural differences between the phases that affect the electronic and phonon properties. The structure motif of the $\alpha$ phase is Ga$_2$ dimers, which form strong covalent bonds leading to bonding and antibonding states that suppress the density of states at the Fermi level. The $\beta$-Ga structure consists of arrays of Ga chains that favor strong coupling between the lattice vibrations and the electronic states near the Fermi level. The increased density of states and strong coupling to the phonons for the $\beta$-Ga chains compared to the $\alpha$ Ga$_2$ dimers enhance superconductivity in the $\beta$-Ga phase.
\end{abstract}
\maketitle
\section{Introduction}
The structural phase diagram of elemental gallium is complex. At ambient pressure, gallium crystallizes in the stable orthorhombic $\alpha$ phase. Several metastable phases, $\beta$, $\gamma$, $\delta$, and $\epsilon$, can also be synthesized under ambient conditions~\cite{alpha-ga, beta-ga,gamma-ga,delta-ga}. Under pressure, Ga exhibits three additional phases, Ga-II, Ga-III, and Ga-V~\cite{Bosio1978, Degtyareva2004}. Most gallium phases ($\alpha$, $\beta$, $\gamma$, $\delta$, Ga-II) undergo superconducting transitions. The superconducting transition temperature, $T_c$, of the stable $\alpha$-Ga phase is about 0.9~K~\cite{alpha-tc}. However, the metastable $\beta$, $\gamma$, and $\delta$ phases have much higher $T_c$'s. Metastable $\beta$-Ga is reported to have a $T_c$ of 5.9 to 6.2~K~\cite{PhysRevB.7.166, PhysRevB.97.184517, FEDER1966611}, while the transition temperatures of $\gamma$ and $\delta$ Ga are 6.9 to 7.6~K~\cite{1047,FEDER1966611} and 7.85~K~\cite{1048}, respectively.

The significant increase in the superconducting transition temperature from 0.9~K in the stable $\alpha$ phase to 5-8~K in the $\beta$, $\gamma$, and $\delta$ phases poses an interesting theoretical question. In the context of the recent discoveries of high-temperature superconductivity in hydrides at megabar pressure~\cite{Flores_Livas2020}, there is great interest in understanding situations where long-lived metastable phases formed at ambient or high pressure might  exhibit a higher $T_c$ than their stable counterparts. The gallium phases represent a unique opportunity to study this problem theoretically to understand the enhancement from the low-$T_c$ stable phase to the higher-$T_c$ metastable phases.

The phase stability of gallium was first studied by Gong {\it et al.}\ and Bernasconi {\it et al.}\ separately in the 1990s~\cite{PhysRevB.43.14277, PhysRevB.52.9988}. Both studies found that the $\alpha$ structure is the most stable phase. Gong {\it et al.}\ pointed out that the structural motif of $\alpha$-Ga consists of Ga$_2$ dimers with strong covalent bonding and significant charge localization inside the dimers. %\cite{C4CP03643C,PhysRevB.46.7319} 
A recent study based on the Wannier functions analysis of the chemical bonding in $\alpha$-Ga also supports covalent bonds within the Ga$_2$ dimers~\cite{doi:10.1080/00268976.2018.1487598}. Gong {\it et al.}\ characterize the metallic $\alpha$-Ga phase with its strong Ga-Ga bonds as ``a metallic molecular crystal''~\cite{PhysRevB.43.14277}.
%Recently, Niu {\it et al} constructed interaction potential by training neural networks based on ab-initio data and found that the metastable $\beta$-Ga is  kinetically more favorable over the $\alpha$ phase above 174 K.\cite{phase} 

Recently, Campanini {\it et al.}\ reported an {\it in situ}  characterization of the $\alpha$ to $\beta$ phase transition using a membrane-based nanocalorimeter~\cite{PhysRevB.97.184517} and found that $\beta$-Ga is a strong coupling type-I superconductor with a $T_c$ of around 6~K~\cite{PhysRevB.97.184517}. Khasanov {\it et al.}\ measured the thermodynamic critical field and specific heat  of the high-pressure Ga-II phase as a function of temperature and demonstrated that it follows the same universal relations as in conventional superconductors~\cite{PhysRevB.101.054504}.
%The magnetoresistance of the single crystal $\alpha$-gallium was found to be large which suggests unusual electronic structure in the $\alpha$ phase of gallium.\cite{chen2018} 
The significant improvement in $T_c$ as a result of the structure phase transition from the stable $\alpha$ structure to the metastable $\beta$ structure motivates us to examine various aspects of the two phases in search of factors that are important for improving $T_c$. In principle it would be interesting to study the even higher-$T_c$ $\gamma$ and $\delta$ phases as well, but  the unit cell of these phases contain 20 and 66 atoms per primitive cell, respectively.  By contrast, the primitive cells of  the $\alpha$ and $\beta$ phases, contain 4 and 2 atoms, respectively.

This paper is structured as follows. Section~\ref{sec:Method} describes the details of the density-functional theory (DFT) calculations and the crystal structure of $\alpha$ and $\beta$ gallium. Section~\ref{sec:Results} compares the electronic structure, charge density distribution, and electron-phonon coupling for the $\alpha$ and $\beta$ phases of gallium. We show that the covalent Ga$_2$ dimers in $\alpha$-Ga suppress the electronic density of states (DOS) at the Fermi level, $E_f$, and that the chains in $\beta$-Ga lead to electronic states near $E_f$ that strongly couple to the phonons, thereby, increasing the superconducting transition temperature. Section~\ref{sec:discussion}, ~\ref{sec:Summary} summarize the results and discuss possible routes for improving the $T_c$ of elemental gallium.

\begin{figure*}[t]
  \includegraphics[width=\textwidth]{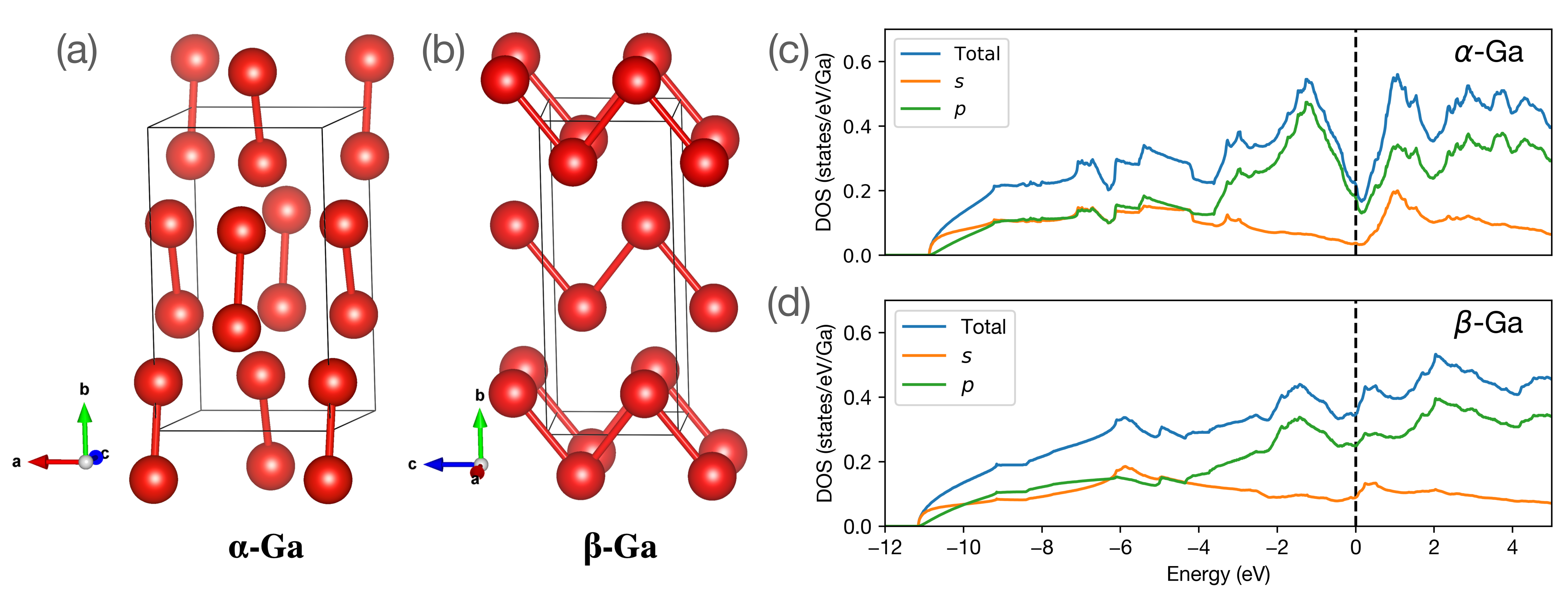}
  \caption{The crystal structure (a,b) and electronic density of states (c,d) of $\alpha$ and $\beta$ Ga. In the $\alpha$ phase (a), Ga atoms form dimers (or dumbbells), which lead to bonding and antibonding states that suppress the density of states at the Fermi level (c). In the $\beta$ phase (b), the Ga atoms form arrays of one-dimensional chains along the $\vec{c}$ lattice vector direction that result in an increased density of states at the Fermi level (d).}
\label{str}
\end{figure*}

\section{Computational method and crystal structure}
\label{sec:Method}

We perform the DFT calculations using the Quantum Espresso code~\cite{QE1,QE2,QE3} with optimized norm-conserving pseudopotential~\cite{PhysRevB.88.085117, SCHLIPF201536} and the PBE  generalized gradient exchange-correlation functional~\cite{Perdew1996}. The plane-wave cutoff is set to 80~Ry and the charge density cutoff to 320~Ry. The $k$-point mesh for the self-consistent calculations is $25\times25\times25$ and the $q$-point mesh for the electron-phonon coupling calculations is $5\times5\times 5$. After computing $\epsilon_k$, $\omega_q$, and $g(k,q)$ on a coarse mesh, we construct the electron and phonon Wannier functions using the EPW code~\cite{NOFFSINGER20102140,PONCE2016116} to interpolate onto fine $k$ and $q$ meshes with $60\times60\times60$ and $30\times30\times30$ points, respectively.

Figure 1 (a) and (b) show the orthorhombic $\alpha$ and the monoclinic $\beta$ gallium phases with space groups $Cmca (64)$ and $C2/c (15)$, respectively. The DFT relaxed lattice parameters for the $\alpha$ phase are $a = 4.55$, $b = 7.71$, and $c=$\SI{4.56}{\AA} and for the $\beta$ are $a=2.79$, $b=8.09$, and $c=$\SI{3.34}{\AA}, and $\beta = 92.2^\circ$, which are within $2\%$ of the experimental values for the lengths of the lattice vectors and within 0.2$^\circ$ for the monoclinic angle~\cite{PhysRevB.43.14277,beta-ga}.

\section{Results}
\label{sec:Results}

\emph{Electronic structure.}
The electron configuration of the gallium atom is $[Ar]3d^{10}4s^24p^1$, with a filled $3d$ shell  that does not participate in the chemical bonding. The $4s$ and $4p$ states hybridize and form bonds between the Ga atoms. Although the valence states of $\alpha$ and $\beta$-Ga consist of the same $4s$ and $4p$ orbitals, the bonding states and their orbital character near the Fermi level differ substantially in the two phases. Figure~\ref{str}(c) shows that in $\alpha$-Ga, states within 2~eV below the Fermi level are predominantly $4p$ states with negligible $4s$ character, while the $4s$ orbital contributes to one-third of the total DOS within 2~eV above the Fermi level. For $\beta$-Ga shown in Fig.~\ref{str} (d), the $4p$ orbitals dominate the DOS within 2~eV above and below the Fermi level, and the variation in DOS near the Fermi level is small. The most important feature in the DOS of $\alpha$-Ga is the V-shaped pseudogap at the Fermi level that substantially reduces the number of states near $E_f$. This pseudogap has been attributed to covalent bonding~\cite{PhysRevB.43.14277, PhysRevB.52.9988}, the crystal structure of the $\alpha$ phase~\cite{PhysRevB.79.045113}, and the interplay between the electronic states and the Brillouin zone geometry~\cite{Zhu_2011}.

\begin{figure}[t]
  \includegraphics[width=\columnwidth]{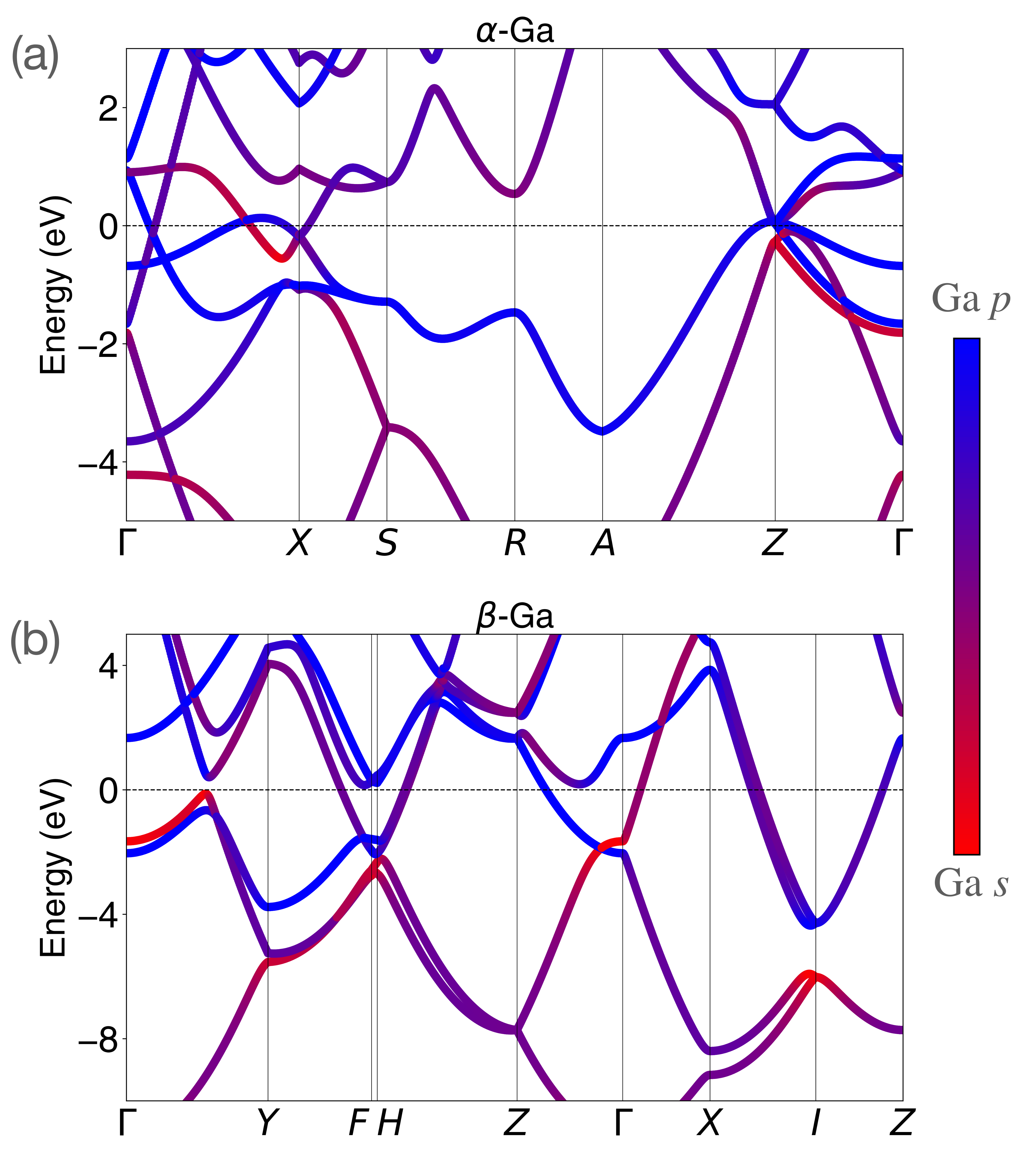}
  \caption{The band structures of $\alpha$ and $\beta$-Ga projected on the $4s$ and $4p$ atomic orbitals show several nearly flat bands in the $\alpha$-Ga and nearly free-electron parabolic bands for $\beta$-Ga.}
  \label{esg}
\end{figure}

Figure~\ref{esg} shows how the band structures of $\alpha$ and $\beta$-Ga differ in several aspects that lead to the observed pseudogap in $\alpha$-Ga and a larger DOS in the $\beta$ phase. The electronic structure of the $\beta$-Ga consists of free-electron-like parabolic bands that span a few eV due to strong intra-chain and inter-chain hoppings.  The band structure of the $\alpha$-Ga, on the other hand, has a few bands that are dispersionless along certain $k$-paths in the Brillouin zone. The flat bands at 1.5~eV below the Fermi level along $X-S-R$  contributes to the DOS peak at $-1.5$~eV. The DOS peak at 1~eV originates from the flat bands along $\Gamma-X-S$. 

\begin{figure}[t]
  \includegraphics[width=\columnwidth]{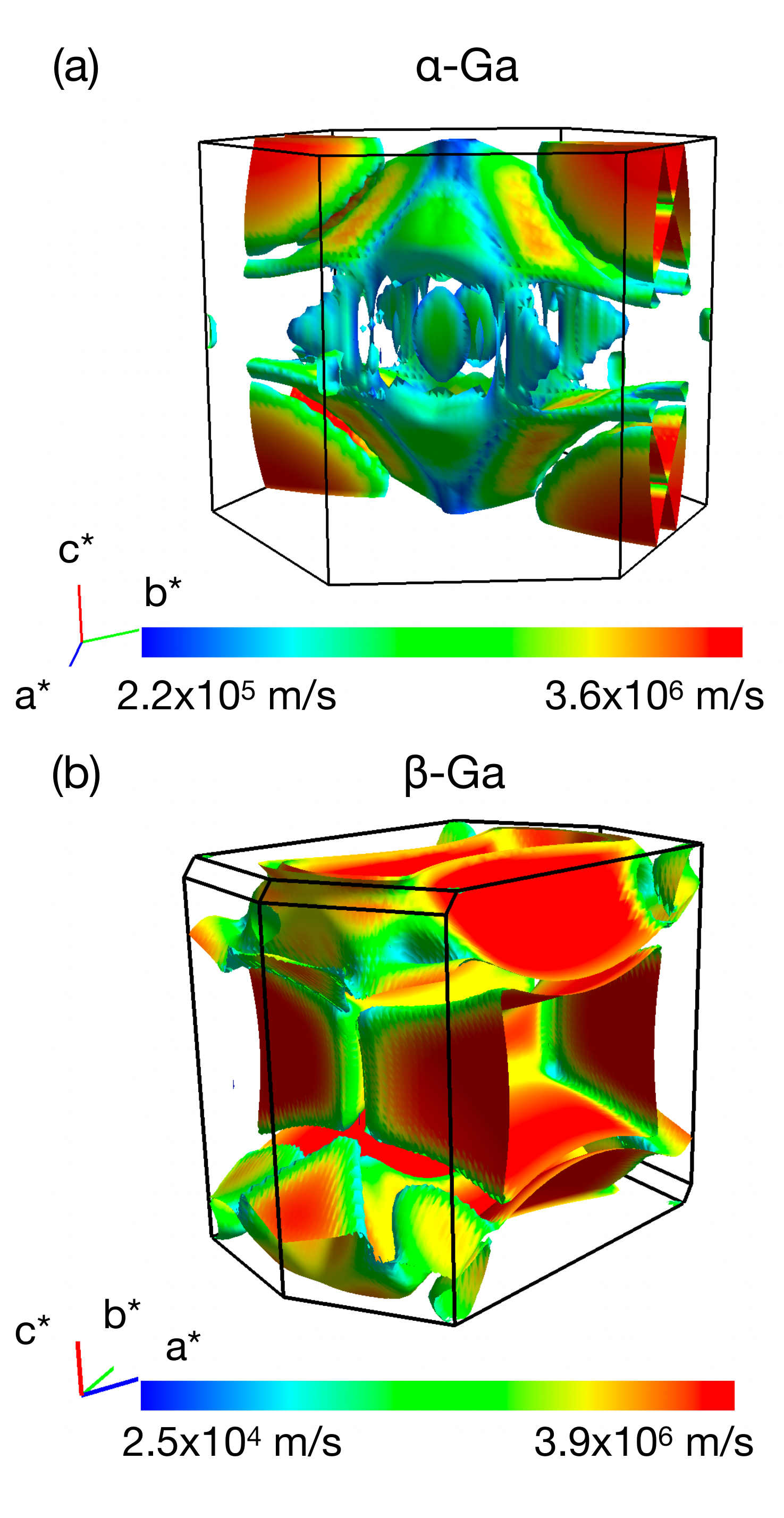}
  \caption{The Fermi surfaces of the (a) $\alpha$ and (b) $\beta$-Ga. Color indicates the Fermi velocity. }
  \label{FS}
\end{figure}

Fig.~\ref{FS} illustrates the Fermi surfaces of the $\alpha$ and $\beta$-Ga. The Fermi surface of $\alpha$-Ga is continuous across the entire Brillouin zone along the  $\vec{a}^\ast+\vec{b}^\ast$ and $\vec{c}^\ast$ directions. Along the $\vec{a}^\ast-\vec{b}^\ast$ direction, the Fermi surfaces is confined to the region from $-0.2 (\vec{a}^\ast-\vec{b}^\ast)$ to $+0.2 (\vec{a}^\ast-\vec{b}^\ast)$, see Fig. S1 in the supplement for detail. 
The Fermi surfaces of $\beta$-Ga extend throughout the Brillouin zone and exhibit flat regions due to its one dimensional chain-like crystal structure, which in part explains the metastability of the $\beta$ phase.

\begin{figure}[t]
  \includegraphics[width=\columnwidth]{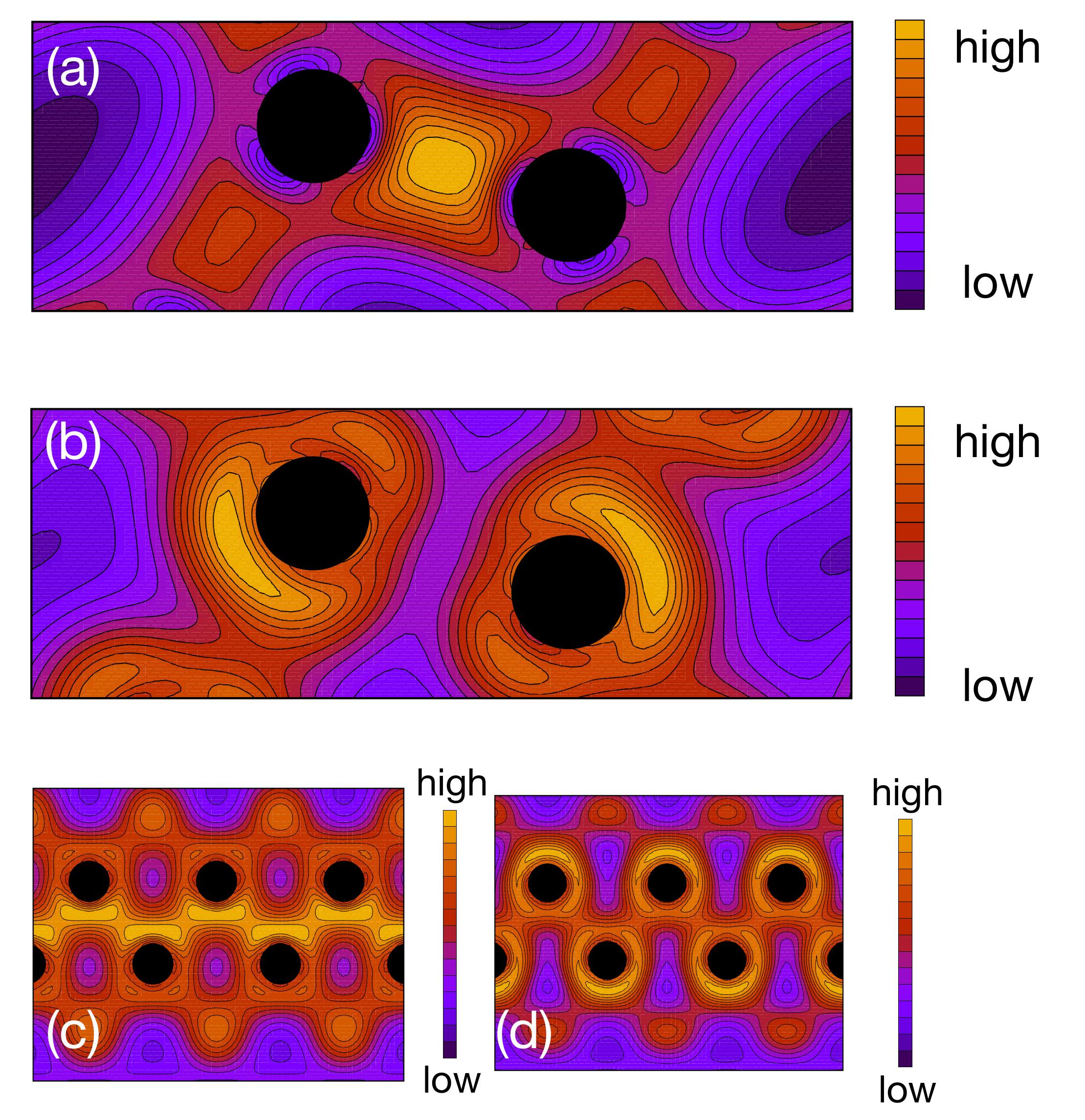}
  \caption{Charge density of $\alpha$ and $\beta$-Ga projected on 1 eV energy windows surrounding the Fermi level. The charge densities for $\alpha$-Ga demonstrate the presence of localized (a) bonding and (b) antibonding states between the Ga$_2$ dimers within 1 eV of the Fermi level.
  For the $\beta$ phase, the charge densities (c) and (d) are delocalized along the Ga-chain, indicating one-dimensional extended states that couple strongly to the lattice.}
  \label{dens}
\end{figure}

\emph{Charge density distributions.}
The pseudogap in the $\alpha$-Ga DOS near the Fermi level strongly suggests the formation of filled bonding and empty antibonding states. Figures~\ref{dens}(a) and (b) show cuts through the charge density of $\alpha$-Ga that include the Ga$_2$ dimer for the energy windows $[-1.5\,\mathrm{eV}, -0.5\,\mathrm{eV}]$ and $[0.5\,\mathrm{eV}, +1.5\,\mathrm{eV}]$, respectively.  The maximum in the charge density between the dimers for the states below the Fermi level and the valley for the states above demonstrate bonding and antibonding states below and above the Fermi level, respectively. Therefore, the formation of strong covalent bonds for the $\alpha$-Ge dimers results in localized states that are shifted away from the Fermi level and, hence, do not participate in the electron-phonon coupling process.

In contrast to $\alpha$-Ga, Figures~\ref{dens} (c) and (d) show that $\beta$-Ga displays a delocalized charge density along the Ga chains. Therefore, the breaking of the Ga dimers liberates the electronic states from the localized molecular bonds, leading to an increased number of states available near the Fermi level that can couple to the lattice vibrations and increase superconductivity.

\begin{figure}[t]
  \includegraphics[width=\columnwidth]{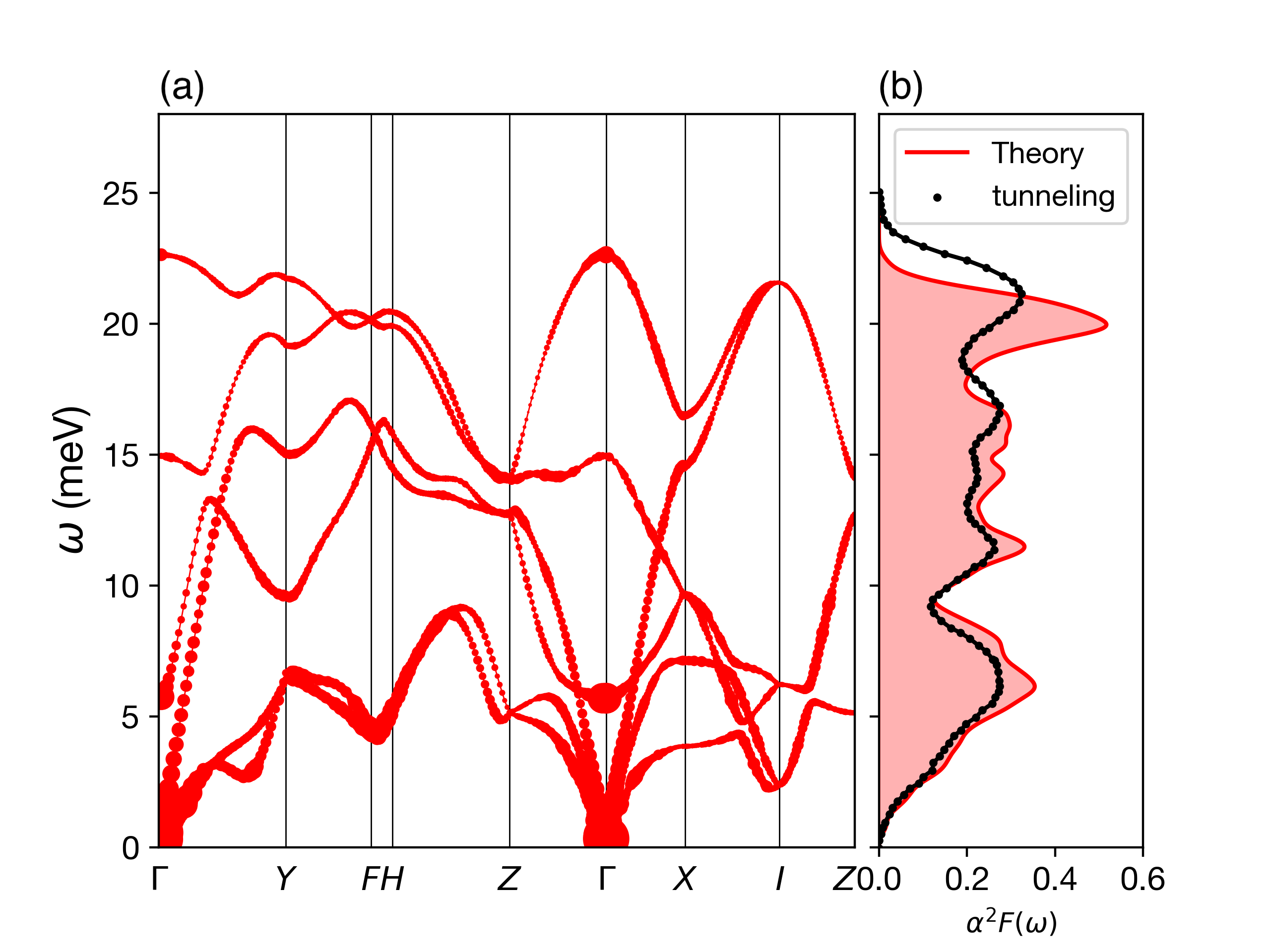}
  \caption{(a) Phonon dispersion of $\beta$-Ga with the size of the data points proportional to $\lambda_q^\nu$ and (b) comparison of $\alpha^2F(\omega)$ obtained from DFT calculations (red) with experimental tunneling data (black) from Ref.~\onlinecite{doi:10.1063/1.1135556}.}
  \label{beta-Ga-a2f}
\end{figure}

\emph{Phonon dispersion and electron-phonon coupling.}
To identify the phonon modes that strongly couple to the electronic states, we plot the phonon dispersion of $\beta$-Ga in Fig.~\ref{beta-Ga-a2f}(a) along the high-symmetry path with the symbol sizes representing the electron-phonon coupling strength $\lambda_{\vec{q}}^\nu$, where $\nu$ denotes the phonon branch index and $\vec{q}$ the wavevector. The two lowest acoustic phonon branches display large electron-phonon coupling $\lambda_{\vec{q}}^\nu$ with noticeable variation along the high-symmetry path. We calculate the mean and standard deviation of $\lambda_{q}^\nu$ for each phonon branch to quantify the strength and anisotropy of the electron-phonon coupling. Table~\ref{tab:var} shows that the average coupling strength of the lowest two phonon branches are 0.36 and 0.25, which accounts for half of the total coupling strength of 1.16. The coupling constants of the  third and fourth branches are 0.16 and 0.15, respectively, about 27\% of the total coupling strength, and the highest two phonon branches contribute 19$\%$ of the total coupling.

\begin{table}[bt]
    \centering
    \caption{The mean and standard deviation of $\lambda_{\vec q}^\nu$ integrated over the entire Brillouin zone for each phonon branch $\nu$ and in total. The electron-phonon coupling of the lowest branch is as large as the sum of the highest three branches.}
    \label{tab:var}
    \begin{ruledtabular}
    \begin{tabular}{c|ccccccc}
     Phonon branch $\nu$   &  1  & 2 & 3 & 4 & 5 & 6 & Total  \\
     \hline
     Mean $\expval{\lambda^{\nu}_{\vec q}}_q$ &0.36 & 0.25 & 0.16 & 0.15  & 0.12 & 0.10 & 1.13\\
     Standard deviation &  0.34 & 0.11 & 0.07 & 0.08 & 0.04 & 0.04 & 0.43 \bf\\
    \end{tabular}
    \end{ruledtabular}
\end{table}

\begin{figure}[!ht]
  \includegraphics[width=1\columnwidth]{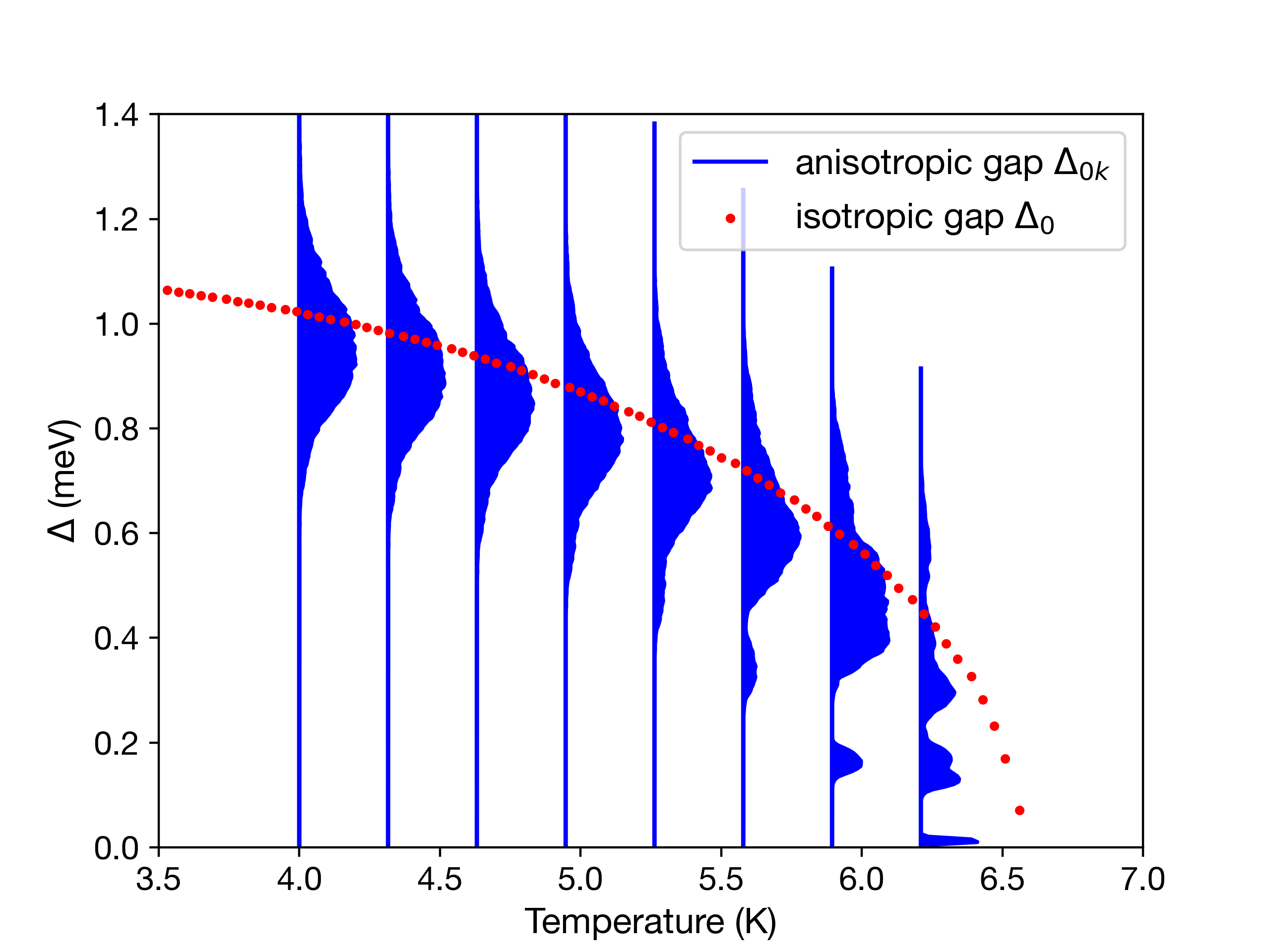}
  \caption{Comparison of the isotropic gap function (red) and the anisotropic gap function (histogram of gap values over the Fermi surface, blue) as a function of  temperature shows that $\beta$-Ga is well described as an isotropic superconductor.}
  \label{fig:gapfunction}
\end{figure}

Figure~\ref{beta-Ga-a2f}(b) compares the calculated electron-phonon coupling $\alpha^2F(\omega)$ obtained with the EPW code for the fine $k$ and $q$-point meshes with the experimental $\alpha^2F(\omega)$ from tunneling experiments~\cite{doi:10.1063/1.1135556} for $\beta$-Ga. The excellent agreement validates the computational methodology. Table~\ref{tab:coupling} summarizes the electron-phonon coupling strength $\lambda$ and the frequency moments $\expval{\omega_2}$ and $\expval{\omega_\mathrm{log}}$ of $\beta$-Ga. Using the Allen-Dynes equation with $\mu^\ast = 0.13$, we estimate a superconducting transition temperature of 6.5~K. Solving the isotropic Eliashberg equations, we obtain essentially the same superconducting transition temperature of 6.6~K. These values closely match the reported experimental $T_c$ of  5.9 to 6.2~K~\cite{PhysRevB.7.166, PhysRevB.97.184517, FEDER1966611}. To characterize the anisotropy of the superconductivity, we also solve the anisotropic Eliashberg equations at several temperatures below $T_c$. Figure~\ref{fig:gapfunction} compares the isotropic and anisotropic gap function as a function of temperature, which demonstrates that $\beta$-Ga is an isotropic superconductor.

\begin{table}[bt]
  \caption{Calculated electron-phonon coupling parameters $\lambda$, $\expval{\omega_2}$, $\expval{\omega_\mathrm{log}}$, and Allen-Dynes critical temperature $T_c^\mathrm{AD}$ for $\mu^\ast=0.13$ of $\alpha$ and $\beta$-Ga. The tunneling data, $\lambda^\mathrm{tunneling}$, is from Ref.~\onlinecite{doi:10.1063/1.1135556}.}
  \label{tab:coupling}
    \begin{ruledtabular}
    \begin{tabular}{c|ccccc}
      &  $\lambda^\mathrm{theory}$ & $\lambda^\mathrm{tunneling}$ & $\expval{\omega_2}$ & $\expval{\omega_\mathrm{log}}$ & $T_c^\mathrm{AD}$ \\
      & & & (meV) & (meV) & (K) \\
     \hline
     $\alpha$-Ga & 0.39 & $-$ & 15.6 & 12.2 & 0.23\\
     $\beta$-Ga & 1.16 & 0.97 & 10.7 & 6.9 &  6.5
  \end{tabular}
  \end{ruledtabular}
\end{table}

\begin{figure}[t]
  \includegraphics[width=0.8\columnwidth]{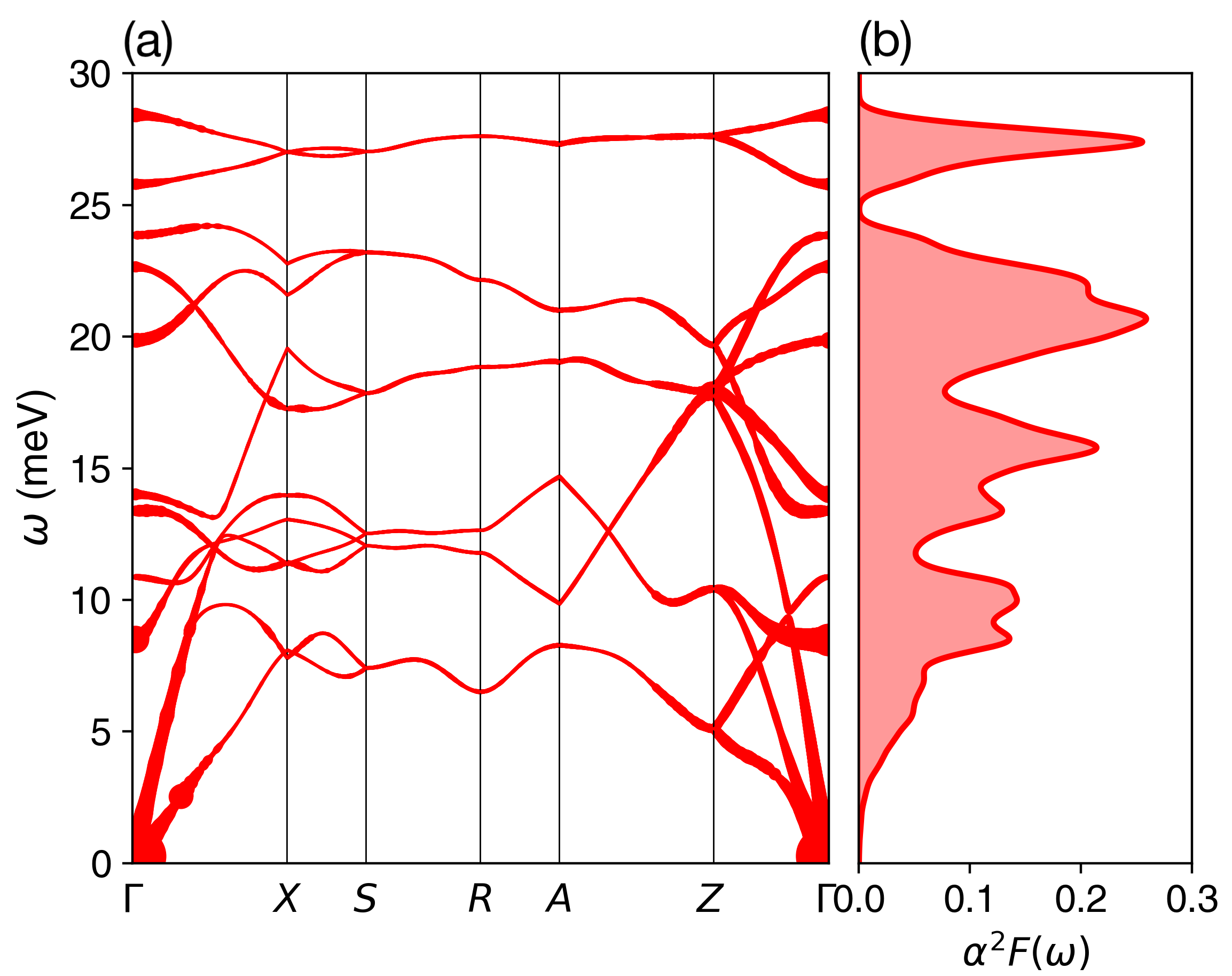}
  \label{alphadis}
  \caption{(a) Phonon dispersion of $\alpha$ gallium along high symmetry path. (b) Eliashberg function of $\alpha$ gallium.  }
  \label{fig:alphaph}
\end{figure}

\begin{figure}[t]
  \includegraphics[width=0.8\columnwidth]{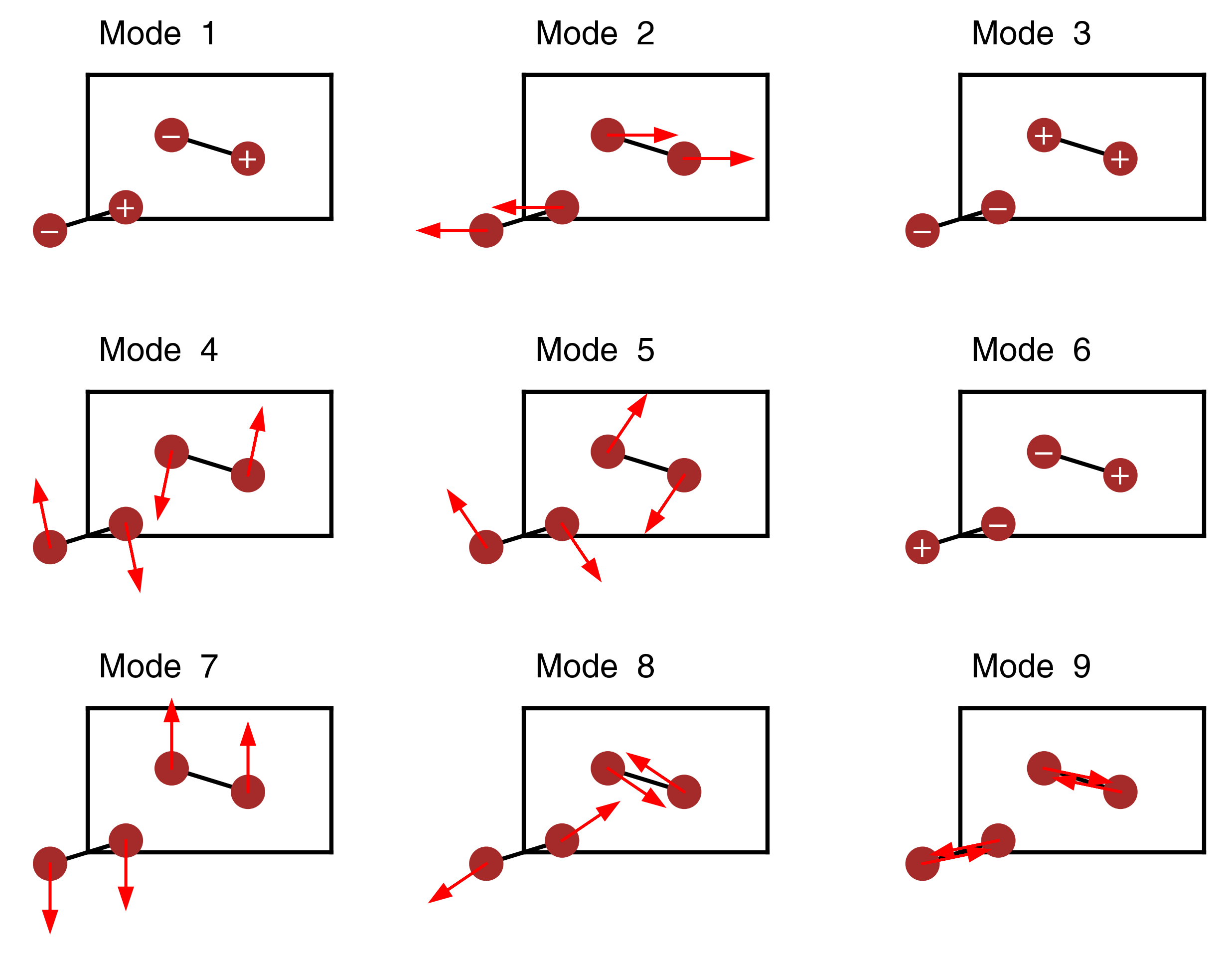}
  \caption{The nine optical phonon modes at the $\Gamma$ point for the $\alpha$-Ga. The plus and minus signs indicate the motion perpendicular to the plane. The two highest frequency modes (mode 11 and 12) are in-phase and out-of-phase Ga-Ga bond stretching modes, respectively. }
  \label{alphaphmod}
\end{figure}

Figures~\ref{fig:alphaph} and \ref{alphaphmod} illustrate the phonon spectrum and the eigenmodes at $\Gamma$ for $\alpha$-Ga, respectively. The primitive cell of the $\alpha$-Ga structure contains four atoms that all lie on the $\vec{b}-\vec{c}$ plane, which represents a mirror plane.
As a result, there are two types of phonon modes at the $\Gamma$ point, in-plane and out-of-plane vibrations, as shown in Fig.~\ref{alphaphmod}. The first and the sixth optical phonon modes comprise of out-of-plane Ga-Ga bond bending, while for the third optical phonon mode, the two Ga-Ga dimers vibrate rigidly into and out of the $\vec{b}-\vec{c}$ plane. The second and the seventh optical phonon modes are the in-plane counterparts of the third optical mode. The eighth and the ninth  phonon modes are bond stretching modes with much higher frequencies than the in-plane bond bending modes (modes 4 and 5). In addition, the phonon dispersions of the highest two phonon branches are gapped from the rest of the phonon spectrum, see Fig.~\ref{fig:alphaph}. The integration of $g(\omega)=\frac{2\alpha^2F(\omega)}{\omega}$ over the frequency range from 25 to 30 meV turns out to be 0.03, about 13$\%$ of the total electron-phonon coupling strength.

\section{Discussion}
\label{sec:discussion}
Several members in the boron group have played an important role in the quest for high-temperature superconductivity. Boron in MgB$_2$, for example, has $p$ orbitals that form delocalized metallic $\sigma$ and $\pi$ bonds. Although $\alpha$-Ga has strong bonds, they are localized and either below or above the Fermi level. Electron or hole doping might shift the Fermi level towards the DOS peaks at +1 or $-1.5$~eV, respectively, which could improve the superconducting transition temperature of the stable $\alpha$ phase. Applying external pressures could be another route for improving the superconducting transition temperatures of $\alpha$ and $\beta$-Ga. Elemental boron {\it per se} is not a superconductor under ambient conditions, although it can transform into a superconductor with a $T_c$ of 4~K at 160~GPa. Further increasing the external pressure to 250~GPa can improve the $T_c$ to 11~K~\cite{Eremets272}. 

Gallium is already a superconductor at ambient pressure. Its $T_c$ ranges from 1 K in the stable $\alpha$ phase to 5-7 K in the metastable $\beta$, $\gamma$, $\delta$ phases. Under external pressure, several new phases are predicted to form~\cite{li2017, doi:10.1063/1.4726256, PhysRevLett.124.145501}. The superconductivity of Ga-II, one of several phases of gallium that emerge under pressure, is measured to be 6.46~K at 3.5~GPa~\cite{PhysRevB.101.054504}. But the superconductivity of gallium under megabar pressures has not been explored systematically. Given that boron, the lightest member of the boron group, undergoes several structure phase transitions before it arrives at its superconducting phase under 160~GPa, new Ga phases with higher $T_c$ might also emerge under extreme pressure.

\section{Summary}
\label{sec:Summary}
We demonstrated that the disparate superconducting properties of gallium's $\alpha$ and $\beta$ phases originate from their difference in bonding using DFT and presented first-principles calculations of the electronic structure, phonon dispersion, electron-phonon coupling, and superconducting properties. For the $\alpha$ phase,  the Ga$_2$ dimers form bonding and antibonding states that result in a $V$-shaped suppression of the DOS at the Fermi level, which significantly reduces the number of states available for coupling to the phonons. The DOS of metastable $\beta$-Ga, on the other hand, is nearly free-electron gas-like due to strong intra- as well as inter-chain hoppings. The charge density confirms the strongly localized bonding and antibonding states below and above the Fermi level in $\alpha$-Ga, while it shows delocalized states near the Fermi level in $\beta$-Ga that are more likely to respond to lattice vibrations. The $T_c$ of $\beta$-Ga is estimated to be 6.5~K, which agrees with the experimental values of 5.9 to 6.2~K~\cite{PhysRevB.7.166, PhysRevB.97.184517, FEDER1966611} and is much higher than the is much higher than the $T_c$ of $\alpha$ Ga. Also, the calculated electron-phonon coupling $\alpha^2F(\omega)$ for $\beta$-Ga closely matches experiment~\cite{doi:10.1063/1.1135556}, validating the computational approach. Therefore, the formation of Ga$_2$ dimers in the $\alpha$ phase suppresses its superconductivity. 

We hope that this understanding of the electronic and phononic structure of the higher-$T_c$ metastable
$\beta$ Ga phase  can pave the way to understanding and designing ambient pressure metastable phases of other
superconductors with higher $T_c$s.

\section{Acknowledgements}
The work was supported by the US Department of Energy Basic Energy Sciences under Contract No. DE-SC-0020385. This research used computational resources provided by the University of Florida Research Computing and the Texas Advanced Computing Center under Contracts TG-DMR050028N. This work used the Extreme Science and Engineering Discovery Environment (XSEDE), which is supported by National Science Foundation grant number ACI-1548562.

\end{document}